\documentclass[a4paper,10pt,twocolumn,showpacs]{revtex4}

\usepackage[latin1]{inputenc}
\usepackage{amsmath}
\usepackage{amssymb}
\usepackage{epsfig}
\usepackage{epstopdf} 
\usepackage{mathrsfs}
\usepackage{color}

\usepackage{siunitx}

\usepackage{graphicx}  
\usepackage{dcolumn}   
\usepackage{bm}        
\usepackage{url}
\usepackage{dsfont}

\newcommand{\e}{\mathrm{e}}

\newcommand{\ket}[1]{|#1\rangle}

\usepackage{color}

\begin{document}

\title{Cooperative emission in quantum plasmonic superradiance}
\author{H. Varguet}
\author{S. Gu\'erin}
\author{H. Jauslin}
\author{G. {Colas des Francs}}\email{gerard.colas-des-francs@u-bourgogne.fr}
\affiliation{Laboratoire Interdisciplinaire Carnot de Bourgogne (ICB), UMR 6303 CNRS, Universit\'e Bourgogne Franche-Comt\'e,
9 Avenue Savary, BP 47870, 21078 Dijon Cedex, France}

\begin{abstract}
Plasmonic superradiance originates from the plasmon mediated strong correlation that builds up between dipolar emitters coupled to a metal nanoparticle. This leads to a fast burst of emission so that plasmonic superradiance constitutes ultrafast and extremely bright optical nanosources of strong interest for integrated quantum nano-optics platforms. We elucidate the superradiance effect by establishing the dynamics of the system, including all features like the orientation of the dipoles, their distance to the particle and the number of active plasmon modes. We determine an optimal configuration for Purcell enhanced superradiance. We also show superradiance blockade at small distances. 
\end{abstract}


\maketitle

\paragraph*{Introduction.}
In a seminal work, Dicke discovered that a set of $N_e$ atoms radiate collectively when they occupy a subwavelength volume. Their emission is much faster ($\tau_{N_e}=\tau_1/N_e$) and stronger ($I_{N_e}=N_e^2I_1$) than for independent atoms. This so-called superradiance originates from spontaneous phase-locking of the atomic dipoles through a same mode and is very similar to the building of cooperative emission in a laser amplifier \cite{Gross-Haroche:1982}.  Superradiant emission produces original states of light with applications such as narrow linewidth lasers \cite{Bohnet-Thompson:2012}  or quantum memories \cite{Afzelius-Gisin-Riedmatten:15,Kim-Oh:17}. Single collective excitation of atoms in  a nanofiber has been demonstrated \cite{Corzo-Laurat:19} and superradiant-like behaviour was suggested in a plasmonics junction \cite{Hess-Hecht:18} or a nanocrystal \cite{JuanPRL:18}, pushing further integration capabilities of quantum technologies. Putsovits and Shahbazyan identifyed plasmon enhanced collective emission for dipoles coupled to a metal nanoparticle (MNP) \cite{Pustovit-Shahbazyan:09}, considering a classical approach which however cannot describe the Dicke cascade at the origin of the cooperative emission. In this communication, taking benefit from recent advances on quantum plasmonics and open quantum systems \cite{vanVlack-Hughes:2012,Tame-Maier:2013,Rousseaux-GCF:2016,DrezetPRA:17,Marquier_Sauvan-Greffet:17,Protsenko:2017,SaezBlaquez-GarciaVidal:17,Asenjo-Kimble:17,Yokoshi:17},  we derive a quantum approach for plasmonic superradiance and discuss the dynamics of cooperative emission with particular attention to the role of the localized surface plasmons (LSP$_n$, where $n$ refer to the mode order).

\begin{table}
\begin{tabular}{l | ccccc}
&\includegraphics[width=3cm]{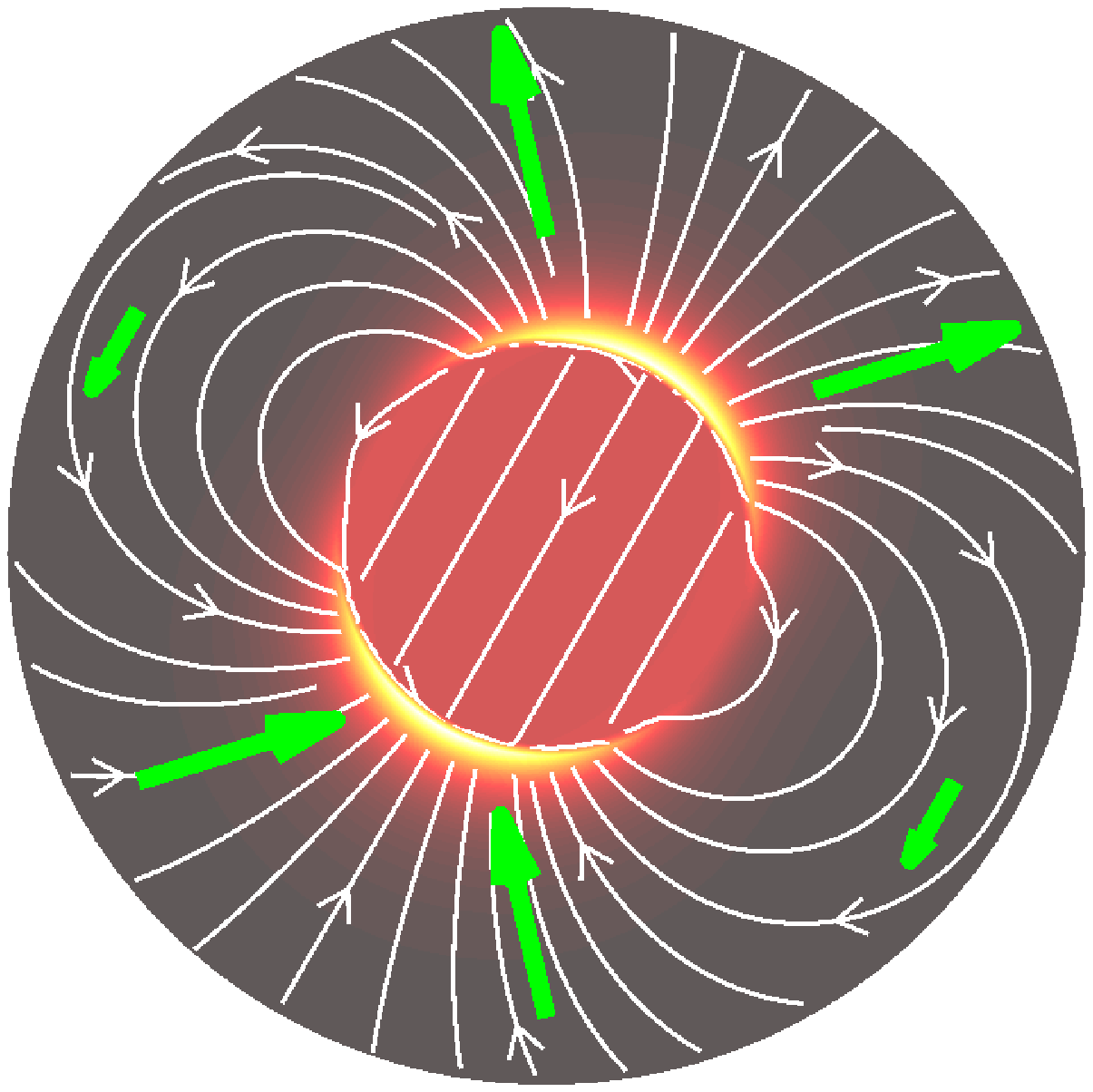}  
&
 & \includegraphics[width=1.2cm]{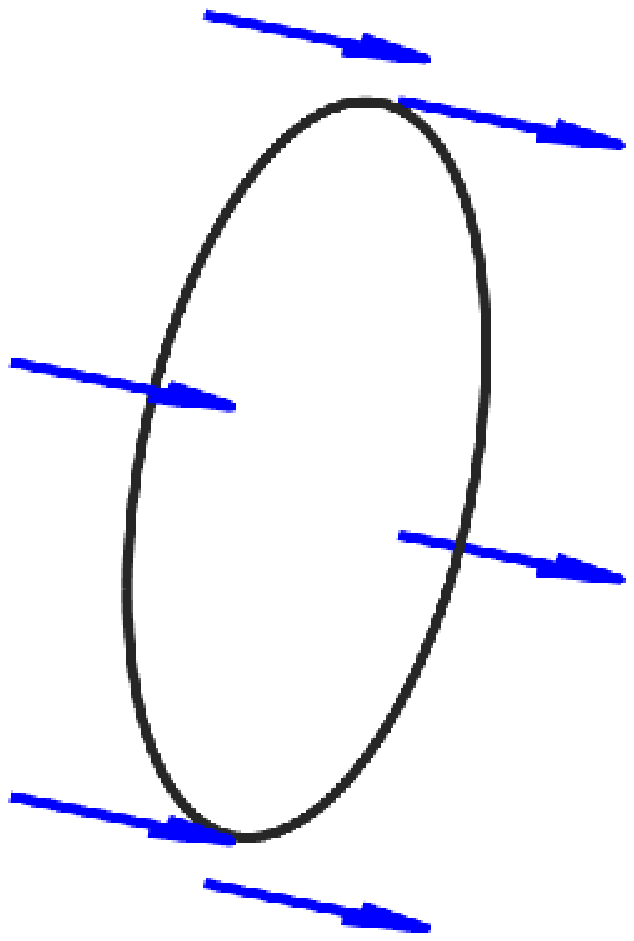}
 &
&\includegraphics[width=1.7cm]{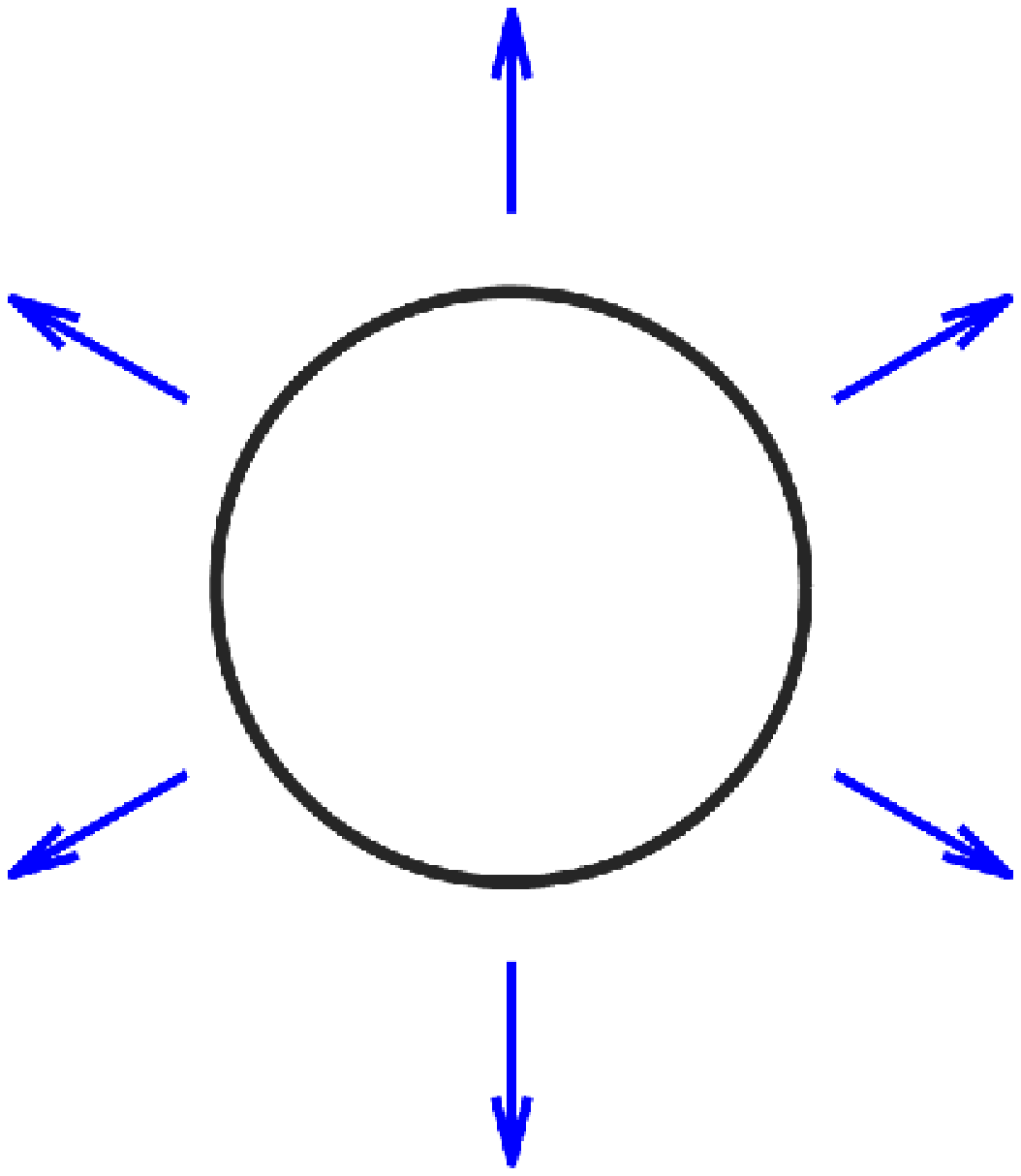} 
\\
\hline \\
$\Gamma_{tot}/\Gamma_0$ &$\SI{325}{}$& &$\SI{93}{}$  & &$\SI{2.7}{}$ \\
$\Gamma_{tot}/\Gamma_1$ &- & &5.74 & &0.03
\end{tabular}
\caption{Bright states with 6 emitters at 20 nm from a 30 nm MNP ($\omega_0=2.77$ eV). The field lines of LSP$_1$ are  superimposed to the brightest configuration.} 
\label{table:Dicke6All}
\end{table}

\paragraph*{Single excitation superradiance} We first consider single excitation superradiance that presents a classical analogue, facilitating the physical representation of the collective process. It reduces to an eigenvalue problem on the dipole moment $\vec d^{(i)}$ of ${N_e}$ emitters  located at $\vec r_i$ \cite{Araujo:16,Pustovit-Shahbazyan:09,Choquette:2010,Fauche-Lalanne:17,Berthelot-Laverdant:18}
\begin{eqnarray}
\left[(i\frac{\Gamma_{tot}}{2}+\Delta_{tot})\mathds{1} -\frac{3}{2k_0^3}\Gamma_0\sum_{j=1}^{N_e}{\bf G}(\vec r_j,\vec r_i,\omega_0)\right] \cdot  \vec d^{(i)}=0 \;, 
\label{eq:ClassicSuper}
\end{eqnarray}
where ${\bf G}$ is the Green tensor in presence of the MNP, $\omega_0$ the angular frequency of emission and $k_0=\omega_0/c$. $\Gamma_0$ is the free-space dipolar decay rate. We assume a Drude behavior $\varepsilon_m(\omega)=\varepsilon_{\infty}-\omega_p^2/(\omega^2+i\gamma_p\omega)$ with $\varepsilon_{\infty}=6$, $\hbar\omega_p=7.90$ eV and $\hbar\gamma_p=51$ meV for silver. 
 \begin{table}[h!]
\begin{tabular}{l | l cccc}
&  \includegraphics[width=1.25cm]{Ag6-Gpar4-2.eps}
 & \includegraphics[width=2.5cm]{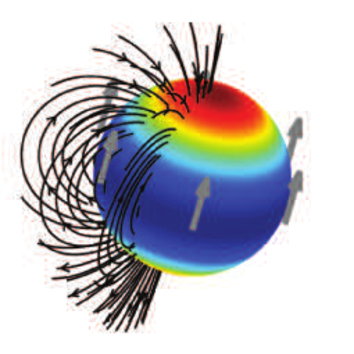} 
 &\includegraphics[width=1.25cm]{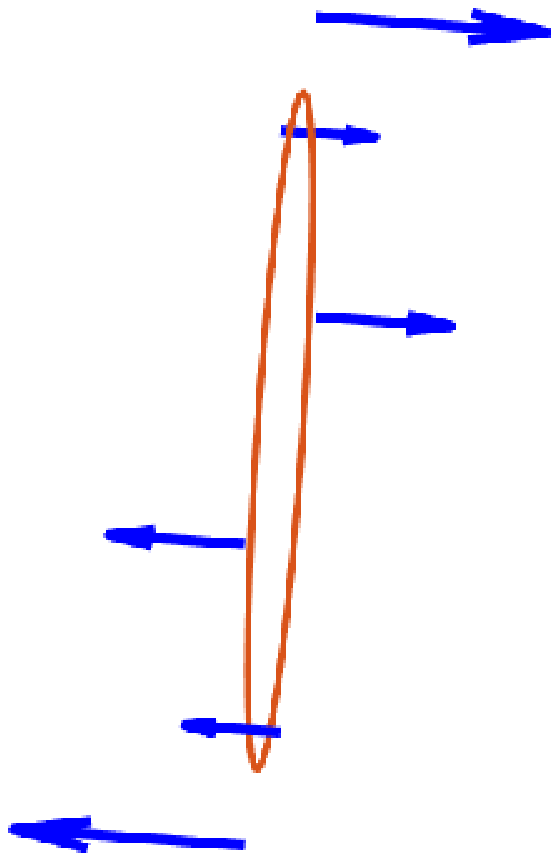}
 &
 &\includegraphics[width=1.25cm]{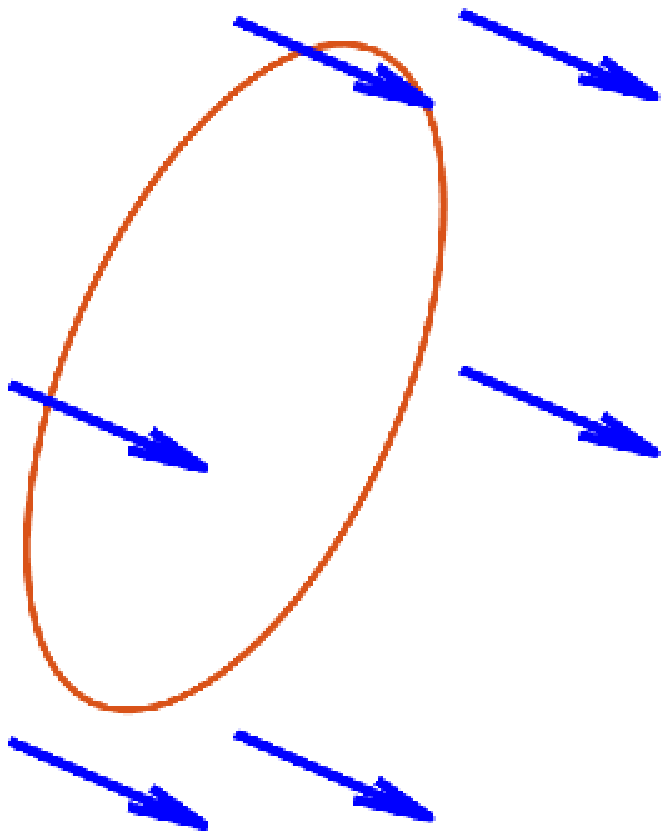} \\
\hline 
&  all LSPs & LSP$_1$ & LSP$_2$ & & LSP$_3$  \\
\hline 
$\Gamma_{tot}/\Gamma_1$& 5.74 & 6 & 3 & & 2.25 
\end{tabular}
\caption{Brightest states maximizing $\Gamma_{tot}/\Gamma_1$ and considering single mode MNP response. The field lines are indicated for LSP$_1$. $\Gamma_1$ refers to single emitter configuration (all LSPs or single mode).}
\label{table:Dicke6}
\end{table}

Typical plasmonic collective states (eigenmodes) are presented on table \ref{table:Dicke6All}. The brightest state ($\Gamma_{tot}/\Gamma_0=325$) is obtained for dipoles that are parallel to the field lines of LSP$_1$, which favors the collective coupling, and most of them (4 out of 6) are almost perpendicular to the MNP surface, corresponding to a strongly enhanced decay rate. We also observe simple configurations maximizing the ratio $\Gamma_{tot}/\Gamma_1$ where $\Gamma_1$ refers to single emitter (azimuthal or radial) coupled to the MNP. For azimuthal emitters $\Gamma_{tot}/\Gamma_1=5.74$, but for radial orientation, $\Gamma_{tot}/\Gamma_1=0.03$ only. In table \ref{table:Dicke6}, we show the role of LSPs for the azimuthal ring arrangement. It presents an ideal superradiant behaviour ($\Gamma_{tot}^1=N_e\Gamma_1^1$) when LSP$_1$ is the only mode involved since all the emitters couple equivalently to LSP$_1$, as displayed by the field lines. This arrangement is also a bright state for LSP$_3$ but not for LSP$_2$ where the dipoles oscillate out of phase. Although the single excitation superradiance can be understood from quantum or classical approaches, only a quantum approach is able to describe the dynamics of the preparation of coherent superradiant states. 

\paragraph*{Quantum master equation.}
The dynamics of $N_e$ emitters coupled to a MNP is governed by an effective Hamiltonian, involving the transition operators $\hat{\sigma}_\pm^{(i)}$ of the emitter $i$, the bosonic operators associated to LSPs \cite{Castellini:18} and fully taking into account the emitter's and LSP's losses \cite{VarguetFano:19}. In the weak coupling regime, the LSPs are practically not populated due to their strong dissipation. The adiabatic elimination therefore permits to transfer the information on the LSPs losses to the effective dynamics of the excited emitters. We obtain the Lindblad master equation for the density operator $\hat \rho$ of the emitters \cite{SuppInfo}
\begin{eqnarray}
\frac{d\hat{\rho}(t)}{dt}=\sum_{j=1}^{N_e}\sum_{k=1}^{N_e}\frac{1}{i\hbar} \left[\hat{H}_{jk} ,\hat{\rho}(t)\right]+\mathcal{D}_{jk}\left[\hat{\rho}(t)\right] \label{rho_eli},
\end{eqnarray}
with 
\begin{subequations}
\begin{eqnarray}
\hat{H} _{jk}&=&-\hbar \Delta_{jk}\,\hat{\sigma}_+^{(k)}\hat{\sigma}_-^{(j)} \;,  
\label{Hjk} \\ 
\nonumber
\mathcal{D}_{jk}\left[\hat{\rho}(t)\right]&=&\Gamma_{jk}\left[\hat{\sigma}_-^{(j)}\hat{\rho}(t)\hat{\sigma}_+^{(k)} \right. 
\\
\label{Djk} 
&& \left. - \frac{1}{2}\left(\hat{\sigma}_+^{(k)}\hat{\sigma}_-^{(j)}\hat{\rho}(t)+\hat{\rho}(t)\hat{\sigma}_+^{(k)}\hat{\sigma}_-^{(j)}\right) \right] 
\end{eqnarray}
\end{subequations}
The parameter $\Gamma_j=\Gamma_{jj}$ ($\Delta_j=\Delta_{jj}$) represents the decay rate (Lamb shift) of the emitter $j$ in presence of the MNP. For $j\ne k$, $\Gamma_{jk}$ and $\Delta_{jk}$ characterize the cooperative decay rate and population transfer. In particular, 
\begin{eqnarray}
\Gamma_{jk}&=&\sum_{n=1}^{N_e}\frac{\gamma_n}{\delta_n^2+\left(\frac{\gamma_n}{2}\right)^2} g_n^{(j)}g_n^{(k)}\mu_n^{(jk)}\label{pertes} \;, 
\end{eqnarray}
where we introduced the coupling strength $g_n^{(j)}$ between the emitter $j$ and the mode LSP$_n$. $\mu_n^{(jk)}$ is the coupling strength between emitters $i$ and $j$ via LSP$_n$ \cite{SuppInfo}. This plays an important role in the emitters' dynamics since, depending on its sign, it can lead to either enhancement or blockade of the cooperative process. 

\paragraph*{Cooperative emission.} In the case of plasmonic Dicke states, the coupling to LSPs strongly depends on the emitter orientation and the number of active LSPs modes. We work in the basis $ \left\{\vert  ee \dots ee \rangle, \mathcal{P}_{N_e}^{N_e-1}\left(\vert \alpha\rangle\right), \dots,\mathcal{P}_{N_e}^{1}\left(\vert \alpha\rangle\right),\vert gg \dots gg\rangle\right\}$, where the permutator $\mathcal{P}_{N_e}^{N_e-l}\left(\vert \alpha\rangle\right)$ gives all the states $\vert \alpha\rangle$ with $N_e-l$ excited emitters. For instance, $\mathcal{P}^{2}_3\left(\vert \alpha\rangle\right)=\left\{\vert eeg\rangle,\vert ege \rangle,\vert gee\rangle\right\}$. 
The collective emission can be written as $I(t)=\hbar \omega_0 W(t)$ with the collective rate
\begin{eqnarray}
W(t)=\left\langle\sum_{i,j=1}^{N_e}\Gamma_{ij}\hat{\sigma}_+^{(i)}\hat{\sigma}_-^{(j)}\right\rangle=Tr\left(\hat \rho \sum_{i,j}\Gamma_{ij}\hat{\sigma}_+^{(i)}\hat{\sigma}_-^{(j)}\right)
\label{def_W}.
\end{eqnarray} 
This expression generalizes the standard definition to the case of non equal rates $\Gamma_{ij}$. It can be separated in two contributions: $W_P$ involving the populations and $W_C$ involving coherences. For instance, in presence of two emitters
\begin{subequations}
\begin{eqnarray}
W_P(t)&=&\left(\Gamma_1+\Gamma_2\right)\langle ee\vert\hat{\rho}(t)\vert ee\rangle 
 \\
\nonumber
&&+\Gamma_1\langle eg\vert\hat{\rho}(t)\vert eg\rangle+\Gamma_2\langle ge\vert\hat{\rho}(t)\vert ge\rangle\label{pop}\\
\nonumber \\
W_C(t)&=&\Gamma_{12}\left[ \langle ge\vert\hat{\rho}(t)\vert eg\rangle+\langle eg\vert\hat{\rho}(t)\vert ge\rangle \right] \;.
\end{eqnarray} 
\end{subequations}

Similar expressions can be derived for an arbitrary number of emitters. For independent emitters ($\mu_{i\ne j}=0$), the decay rates $\Gamma_{ij}$ cancel for $i\ne j$. The emission rate reduces to incoherent emission $W_P$ of the independent emitters. The second term $W_C$ describes the  collective behaviour of the ensemble of emitters when $\mu_{ij} \ne 0$. Therefore, the cooperative behaviour originates from the correlation between the states of same excitation $\mathcal{P}_{N_e}^{N_e-l}(\ket{\alpha})$ in full analogy with free-space superradiance. 

Finally, we emphasize that the collective rate $W(t)$ includes both radiation in the far-field and non radiative transfer to the absorbing MNP. Since the cooperative decay rate can be written equivalently as $\Gamma_{jk}=2\omega_0^2/(\hbar\varepsilon_0 c^2)\mathfrak{Im}\left[\mathbf{d}^{(j)}\cdot \mathbf{G}(\mathbf{r}_j,\mathbf{r}_k,\omega_0)\cdot \mathbf{d}^{(k)}\right]$ \cite{SuppInfo},  one can isolate the radiative rates $\Gamma_{ij}^{rad}$ in the quasi-static approximation \cite{Pustovit-Shahbazyan:09}, and the far-field radiated emission obeys 
$I_{rad}(t)=\hbar \omega_0 W_{rad}(t)$ with
\begin{eqnarray}
W_{rad}(t)=\left\langle\sum_{i,j=1}^{N_e}\Gamma_{ij}^{rad}\hat{\sigma}_+^{(i)}\hat{\sigma}_-^{(j)}\right\rangle \,.
\end{eqnarray}
\begin{figure}
\includegraphics[width=0.45\textwidth]{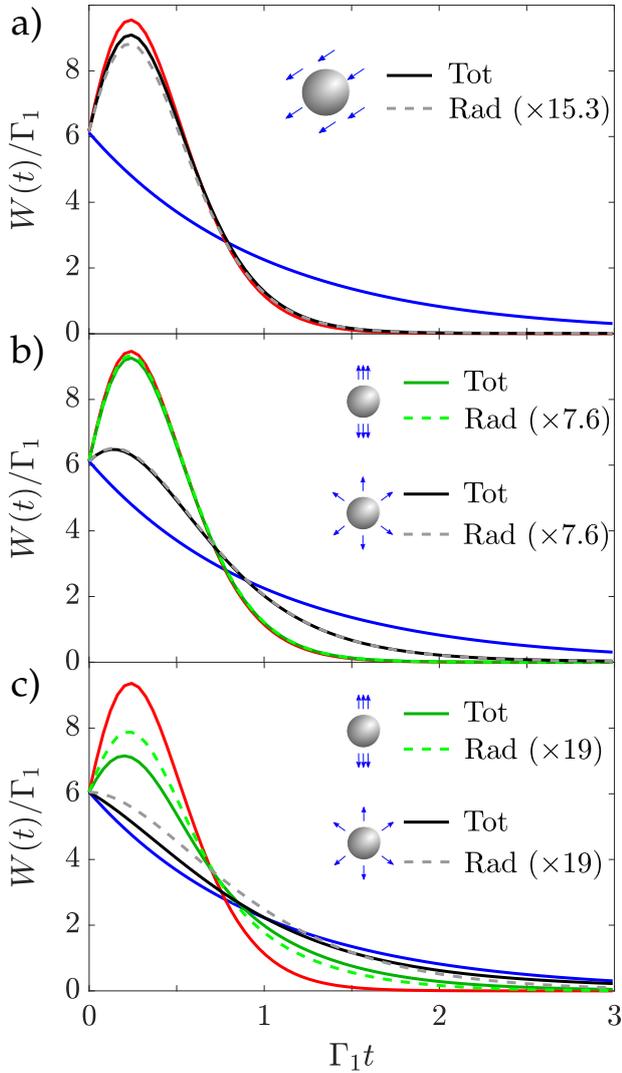} 
\caption{Normalized collective rate $W(t)/\Gamma_1$ for 6 emitters ($\omega_0=\omega_1=2.77$ eV). a) Azimuthal orientation, $h=20$ nm, b) radial orientation, $h=20$ nm,  c) radial, $h=2$ nm. Red curves: ideal superradiance, blue curves: incoherent emission. Green and black curves correspond to emitters at the poles or homogeneously distributed, respectively. Dashed lines refer to the radiative collective rate $W_{rad}$.}
\label{It} 
\end{figure}
The master equation \eqref{rho_eli} is solved numerically following Ref. \cite{Eleuch-Guerin-Jauslin:12,NavarreteBenlloch:15} assuming the initial state $\vert \psi(0)=\vert e\ldots e\rangle$. Figure \ref{It} represents the dynamics of correlated emission. We consider three configurations (black lines): azimutal emitters (perpendicular to the equator) at distance $h=20$ nm in Fig. \ref{It}a and radial emitters at $h=20$ nm (Fig. \ref{It}b), or  $h=2$ nm (Fig. \ref{It}c), and their comparison with two limit cases: ideal superradiance (emitters at the same position, red lines) and incoherent emission ($W_P$, blue lines). 
In case of azimutal orientation and for $h=20$ nm (Fig. \ref{It}a), all the emitters equally couple to LSP$_1$ so that close to ideal superradiance is observed for both total and radiative emissions with a collective quantum yield $\eta=W_{rad}/W=1/15.3=7\%$. For radial emitters, we still observe the burst of emission but less pronounced (Fig. \ref{It}b). At short distances ($h=2$ nm,  Fig. \ref{It}c), almost incoherent emission occurs. The cooperative behaviour is partially recovered for $20$ nm ($\eta=13\%$) and $h=2$ nm ($\eta=5\%$) if the emitters are at the poles (green curves in Fig. \ref{It}b,c), a configuration achievable by nanoscale photopolymerization \cite{Zhou-GCF-Bachelot:2015}. 

\paragraph*{Plasmonic Dicke states (ideal superradiance).} When all the emitters are located at the same position, $\Gamma_{ij}=\Gamma_1$, $\Delta_{ij}=\Delta_1$ so that we can work with the Dicke ladder ($J=N_e/2$)
\begin{eqnarray}
\vert J,M\rangle=\sqrt{\frac{\left(J+M\right)!}{N_e!\left(J-M\right)!}}\hat{J}_-^{J-M}\vert ee\dots  e\rangle \,, 
\end{eqnarray}
where we have introduced the collective spin operator $\hat{J}_{\pm}=\sum_{i=1}^{N_e}\hat{\sigma}_{\pm}^{(i)}$. The state $\vert J,J\rangle=\vert ee\dots e\rangle$ has all the emitters in their excited state and the symmetrized Dicke state $\vert J,M\rangle$ is a superposition of the states with $J+M$ excited emitters. 
The master equation (\ref{rho_eli}) simplifies to 
\begin{eqnarray}
\label{equation_JM}
\frac{d\rho_M(t)}{dt}&=&\Gamma_{M+1}~\rho_{M+1}(t)-\Gamma_M~\rho_{M}(t) \,,
\nonumber
\end{eqnarray}
where  $\rho_{M}(t)=\langle J,M\vert\hat{\rho}(t)\vert J,M\rangle$ and $\Gamma_M=(J+M)(J-M+1)\Gamma_1$ are the population and the collective rate of the state $\ket{J,M}$, respectively. The dynamics is the exact analogue of free-space superradiance except that the decay rate is replaced by its value in presence of the MNP, and superradiance is enhanced by the Purcell factor $\Gamma_1/\Gamma_0$. Finally, the plasmonic superradiance originates from the cascade along the Dicke states ladder. Beginning with the initial condition $\vert \psi(t=0)\rangle=\vert ee\dots e\rangle=\vert  J,J\rangle$, the system successively goes through the Dicke states $\vert J,M\rangle$ with $M=J-1,J-2,\dots$, down to the final ground state $\vert J,-J\rangle=\vert gg\dots g\rangle$. The decay rate starts from $\Gamma_{M}=N_e\Gamma_1$ for $M=J$, increases up to $\Gamma_{M} \approx N_e^2 \Gamma_1$ for $M=0, \pm 1/2$ (depending on the parity of $J$) and then decreases down to $\Gamma_{M}=0$ for the final ground state. The build-up of  this cooperative behaviour is shown in Fig. \ref{It} (red curves). Since the direct dipole-dipole coupling is negligible compared to the LSP mediated dipole-dipole coupling, we avoid the van der Waals dephasing observed for free space configurations \cite{Gross-Haroche:1982}.
\begin{figure}
\includegraphics[width=0.5\textwidth]{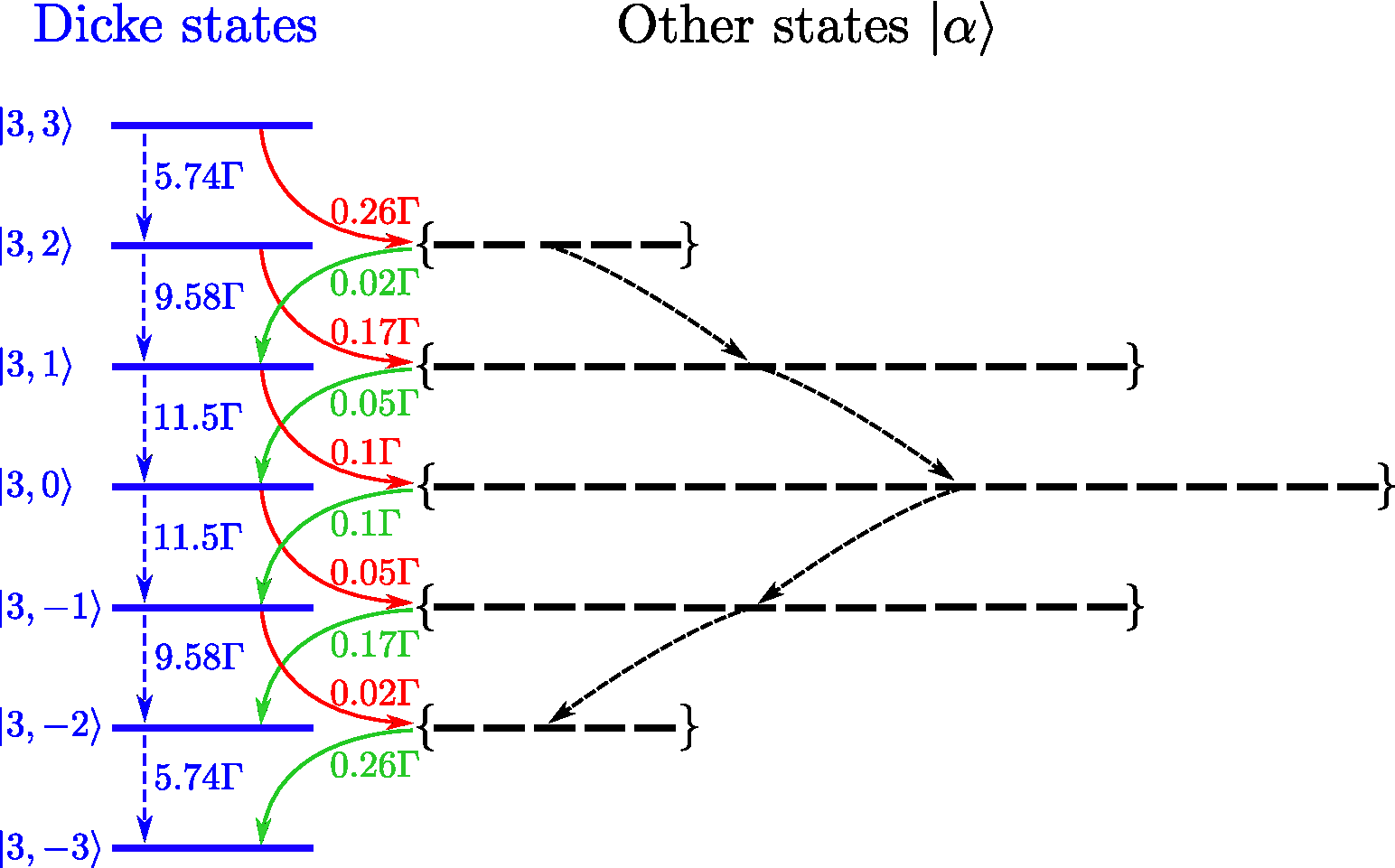} 
\caption{Superradiance cascade along extended Dicke basis for 6 azimuthal emitters homogeneously distributed (same as black curve in Fig. \ref{It}a).}
\label{fig:ExtDike6} 
\end{figure}
\paragraph*{Extended Dicke basis} Moving away from the ideal configuration, it is necessary to generalize the superradiance ladder to describe the full dynamics \cite{Damanet:16}. In case of small deviation from ideal superradiance, it is worth working in an extended Dicke basis including the Dicke states. The decay rates along the Dicke ladder are calculated from the expression 
$\Gamma_{M\rightarrow M'}=\langle\widetilde{M'} \vert \mathcal{D}_{\vert \rho\rangle}\vert  \widetilde{M} \rangle$ 
where $\vert  \widetilde{M} \rangle$ is the vector representation of the projector $\vert  J,M\rangle \langle   J,M \vert$ on the Dicke states and $\mathcal{D}_{\vert \rho\rangle}$ is the associated representation of the dissipator $\mathcal{D}$  \cite{NavarreteBenlloch:15,SuppInfo}. The parasitic transitions to all other states $\vert \alpha \rangle$ follows  $\gamma_{M\rightarrow \alpha}=\Gamma_{M}-\Gamma_{M\rightarrow M'}$ where $\Gamma_{M}=\langle J,M \vert \sum_{i,j=1}^{N_e}\Gamma_{ij}\hat{\sigma}_+^{(i)}\hat{\sigma}_-^{(j)} \vert  J,M \rangle$ is the decay rate of the Dicke state $\vert  J,M\rangle$. 
Similar formulas are used to estimate the input rates $\gamma_{\alpha \rightarrow M'}$ from states $\vert \alpha \rangle$.  Namely, $\gamma_{\alpha \rightarrow M'}=\langle J,M' \vert \sum_{i,j=1}^{N_e}\Gamma_{ij}\hat{\sigma}_-^{(i)}\hat{\sigma}_+^{(j)} \vert J,M' \rangle -\Gamma_{M\rightarrow M'}$. 
Figure \ref{fig:ExtDike6} presents the superradiance cascade along the extended Dicke basis. The decay rates along the Dicke ladder ($\Gamma/\Gamma_1 : 5.75\rightarrow 9.58 \rightarrow 11.5 \rightarrow 11.5 \rightarrow 9.58 \rightarrow 5.75$) closely follow the ideal values ($\Gamma/\Gamma_1 : 6\rightarrow 10 \rightarrow 12 \rightarrow 12 \rightarrow 10 \rightarrow 6$) for azimuthal emitters so that one still observes a burst of emission in Fig. \ref{It}a (black solid line). Parasitic transitions outside the Dicke ladder slightly degrade the collective emission. 

\paragraph*{Role of LSPs.} We discuss the role of LSPs on the superradiance emission considering two emitters located at the same distance to the MNP (but not the same position). We work in the Dicke basis $\ket{ee}, \ket{S}, \ket{A}, \ket{gg}$ with the symmetric and antisymmetric states
\begin{eqnarray}
\vert S\rangle=\frac{1}{\sqrt{2}}\left(\vert ge\rangle+\vert eg\rangle\right) \;, \;
\vert A\rangle=\frac{1}{\sqrt{2}}\left(\vert ge\rangle-\vert eg\rangle\right) \,.
\end{eqnarray}
Their populations dynamics are given by
\begin{subequations}
\begin{eqnarray}
\partial_t\rho_{ee}(t)&=&-2\Gamma_1\rho_{ee}(t),\\
\partial_t\rho_{S,A}(t)&=&\Gamma_{S,A}\rho_{ee}(t)-\Gamma_{S,A}\rho_{S,A}(t),\\
\partial_t\rho_{gg}(t)&=&\Gamma_S\rho_{S}(t)+\Gamma_A\rho_{A}(t) \;,
\end{eqnarray}
\end{subequations}
\begin{figure}
\includegraphics[width=0.32\textwidth]{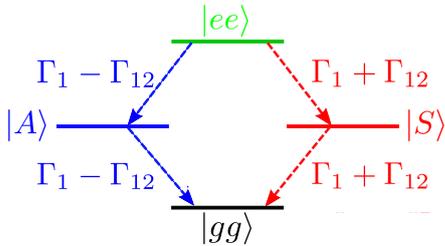}
\caption{Dicke ladder for two emitters-MNP configuration.\label{scheme12}}
\end{figure}
as displayed on Fig. \ref{scheme12}. The population of the excited state $\ket{ee}$ decays exponentially with the rate $\Gamma_{ee}=2\Gamma_{1}$ that does not depend on the angle between the emitters. The symmetric and anti-symmetric states are populated from $\ket{ee}$ with the rates $\Gamma_{S,A}=\left(\Gamma_1\pm \Gamma_{12}\right)$ and relax towards the ground state with the same rates. Finally, the collective decay  rate is
\begin{eqnarray}
\label{taux_photon}
W(t)&=&2\Gamma_1\rho_{ee}(t)+\Gamma_S\rho_S(t)+\Gamma_A\rho_A(t) \;.
\end{eqnarray}

\begin{figure}[h!]
\includegraphics[width=8cm]{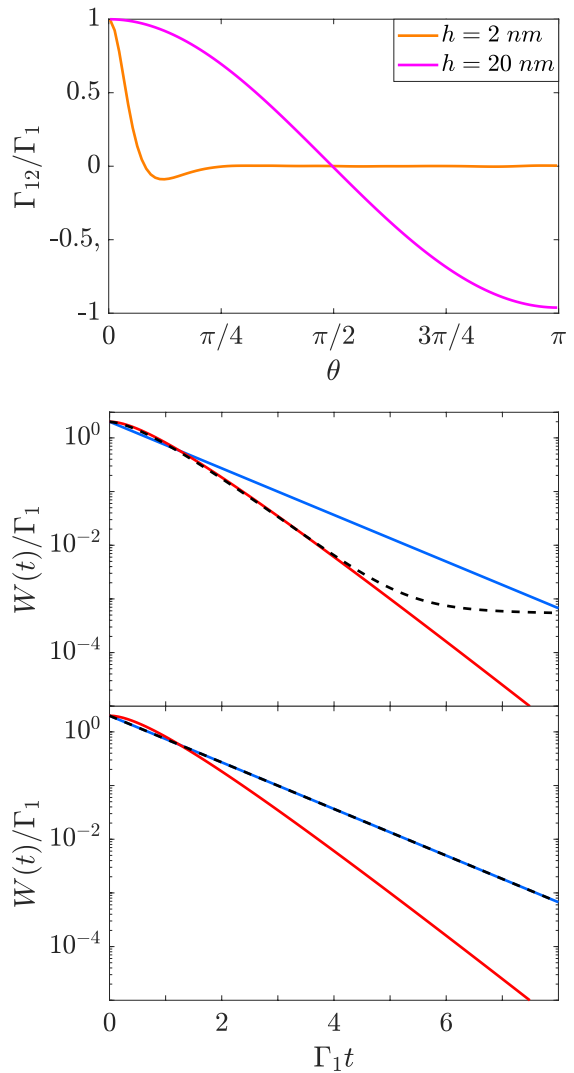}
\caption{a) Two emitters cooperative decay rate $\Gamma_{12}$ as a function of their angular separation $\theta$, for two distances $h=2$ or $20$ nm. 
b, c) Collective rate $W(t)$ for 2 emitters . (b) $h=20$ nm ($\omega_0=2.771$ eV) and (c) $h=2$ nm ($\omega_0=2.964$ eV). In (b,c), red curves correspond to ideal superradiance, blue curves to incoherent emission and black curves to the emitter located at the poles.\label{It_log}}
\end{figure}
The cooperative relaxation strongly depends on $\Gamma_{12}$, and thus on the emitter's positions. Figure \ref{It_log}a) shows the cooperative rate $\Gamma_{12}(\theta)$. For two emitters at the same position, $\Gamma_{12}(0)=\Gamma_1$, and we recover the ideal superradiant configuration. The bright superradiant state $\vert S\rangle$ decays with the rate $\Gamma_S=N_e\Gamma_1$ ($N_e=2$) and the dark subradiant state is $\vert A\rangle$ which is not populated and presents a zero decay rate $\Gamma_A=0$. For emitters located at the poles ($\theta=\pi$) the collective behaviour depends on the distance to the MNP. For large separation distances, only the dipolar LSP$_1$ mode significantly contributes to the emitters-MNP coupling and $\Gamma_{12}(\pi)\approx-\Gamma_1$.  Superradiant and subradiant states are exchanged ($\vert A\rangle$ and $\ket{S}$, respectively), but the collective dynamics (black curve, Fig. \ref{It_log}b) is close to the ideal case. At smaller distances, the cooperative behaviour is inhibited ($\Gamma_{12}\approx 0$) in presence of high order LSPs because of destructive superposition of their contribution \cite{Rousseaux-GCF:2016,Castellini:18}. This results in a superradiance blockade and the dynamics closely follows an incoherent process (Fig. \ref{It_log}c). 

 \begin{table}
\begin{tabular}{l | cccc}
&
\includegraphics[width=1.75cm]{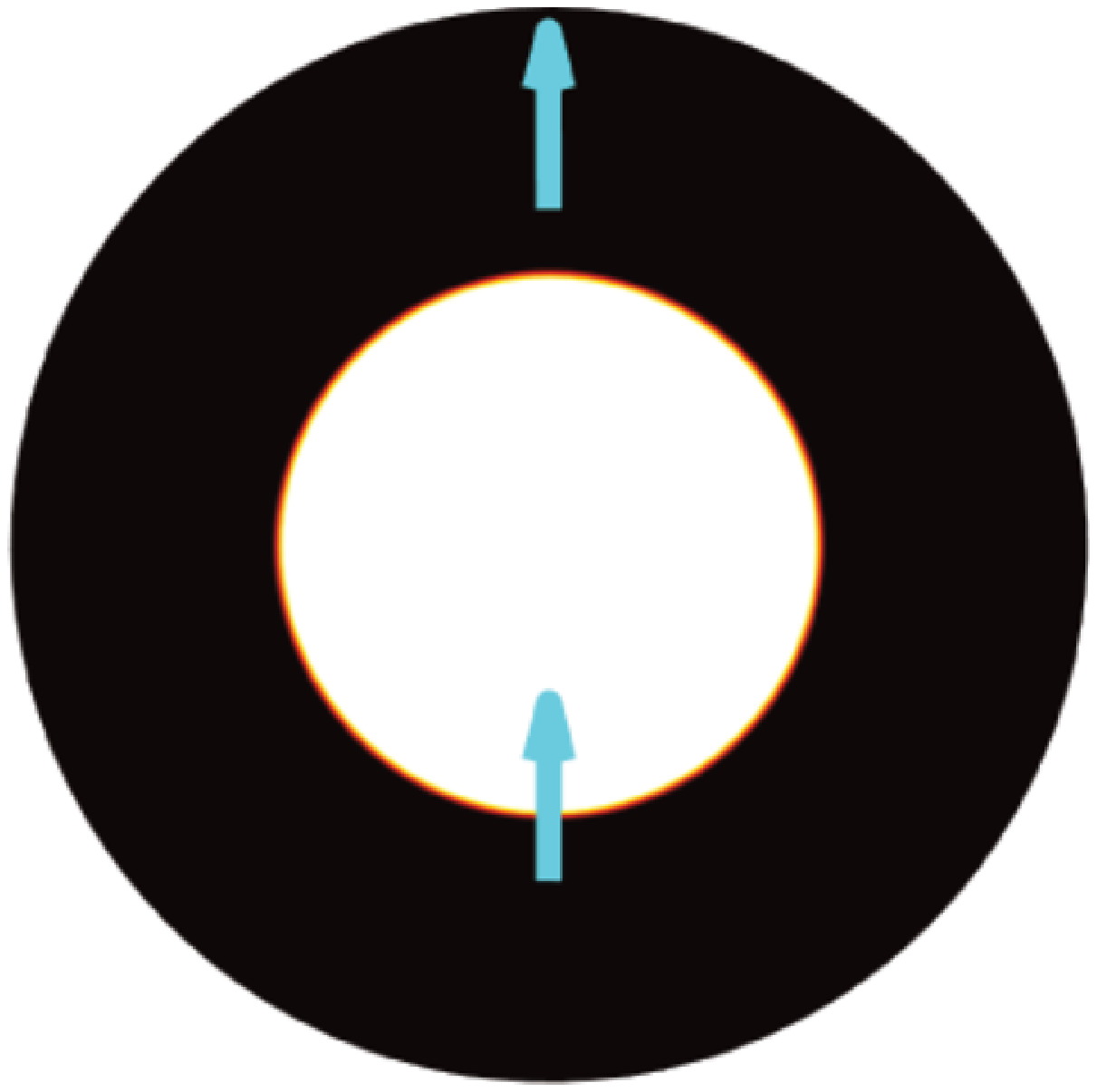} &
\includegraphics[width=1.75cm]{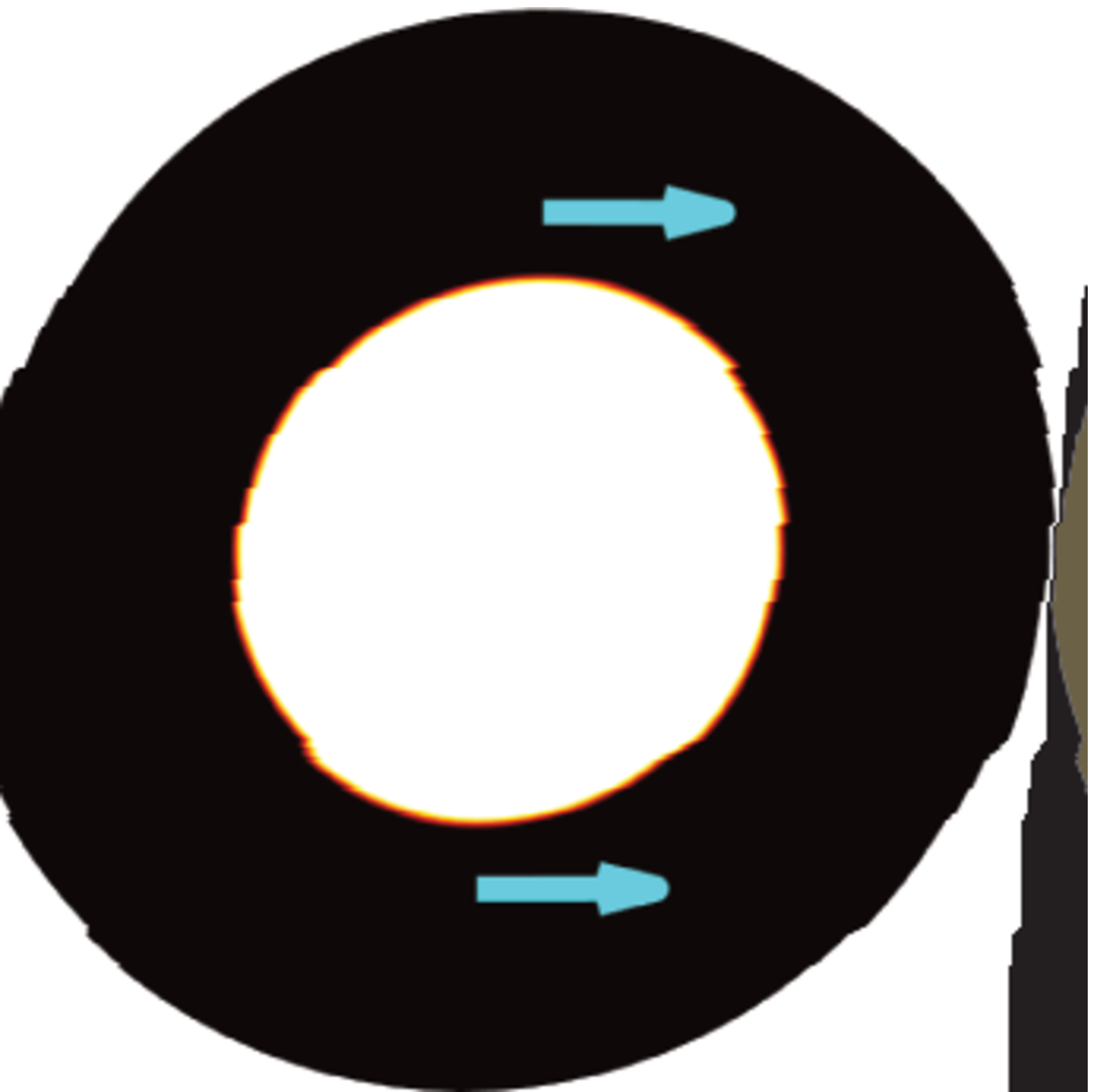} &
\includegraphics[width=1.75cm]{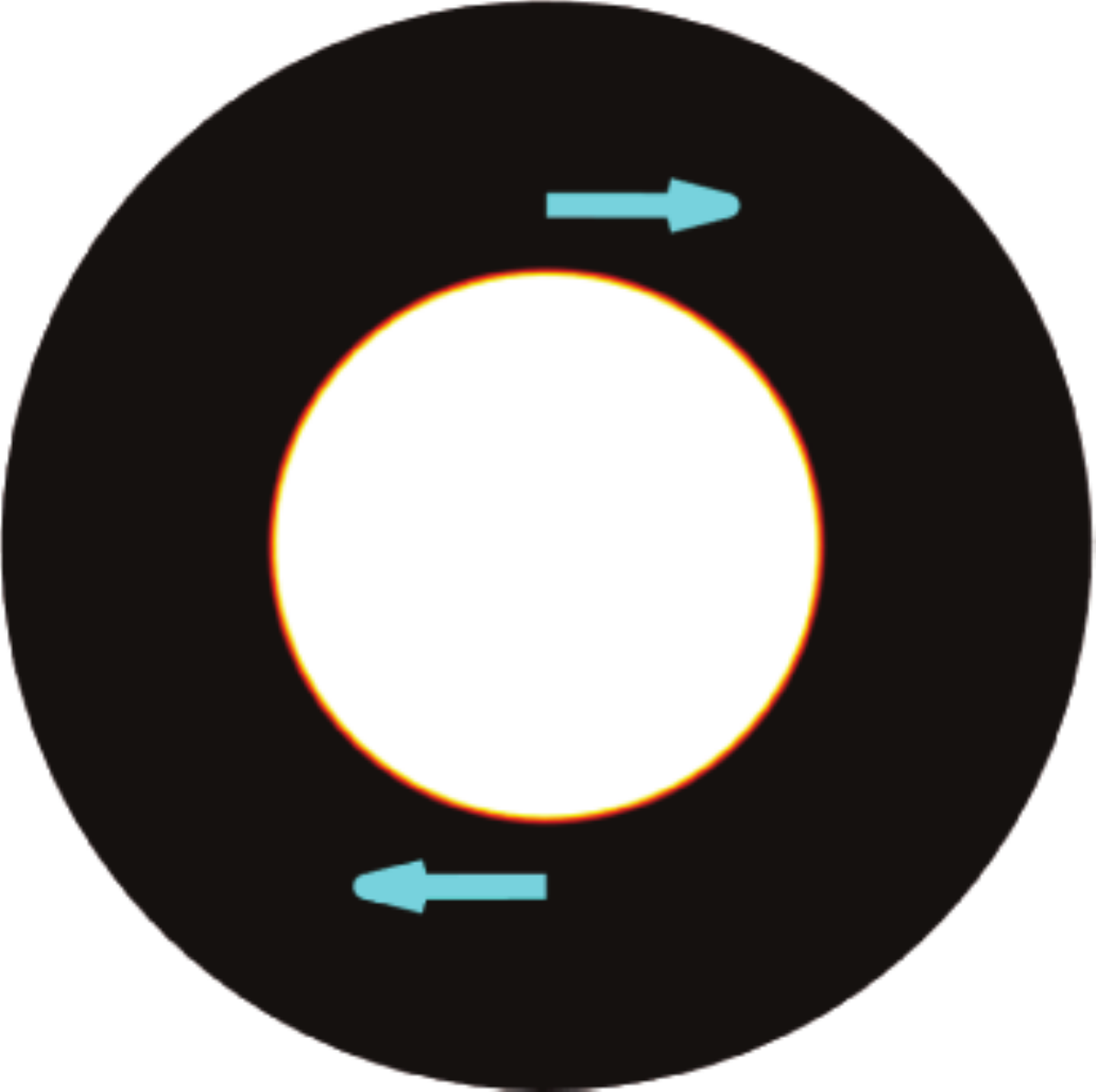} &
\includegraphics[width=1.75cm]{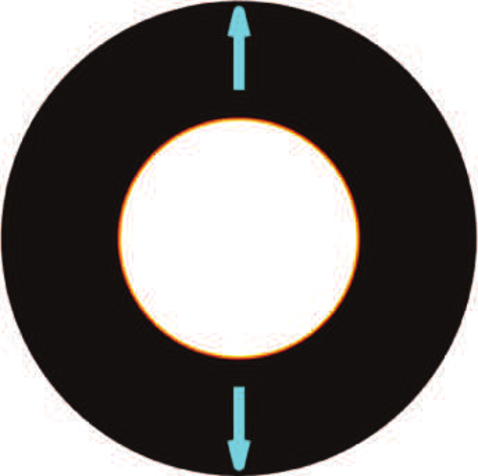} \\
\hline \hline 
$\frac{\Gamma_2}{\Gamma_1}$ &1.96 & 1.92 & 0.08 & 0.04 
\end{tabular}
\caption{Plasmonic super/sub-radiant states with 2 emitters ($h=20$ nm, $\omega_0=2.771$ eV). The decay rate is normalized with respect to a single emitter decay rate for the same configuration (radial or azimuthal).} 
\label{table:Dicke2}
\end{table}

\begin{table}
\begin{tabular}{l | ccc}
 & \includegraphics[width=2.25cm]{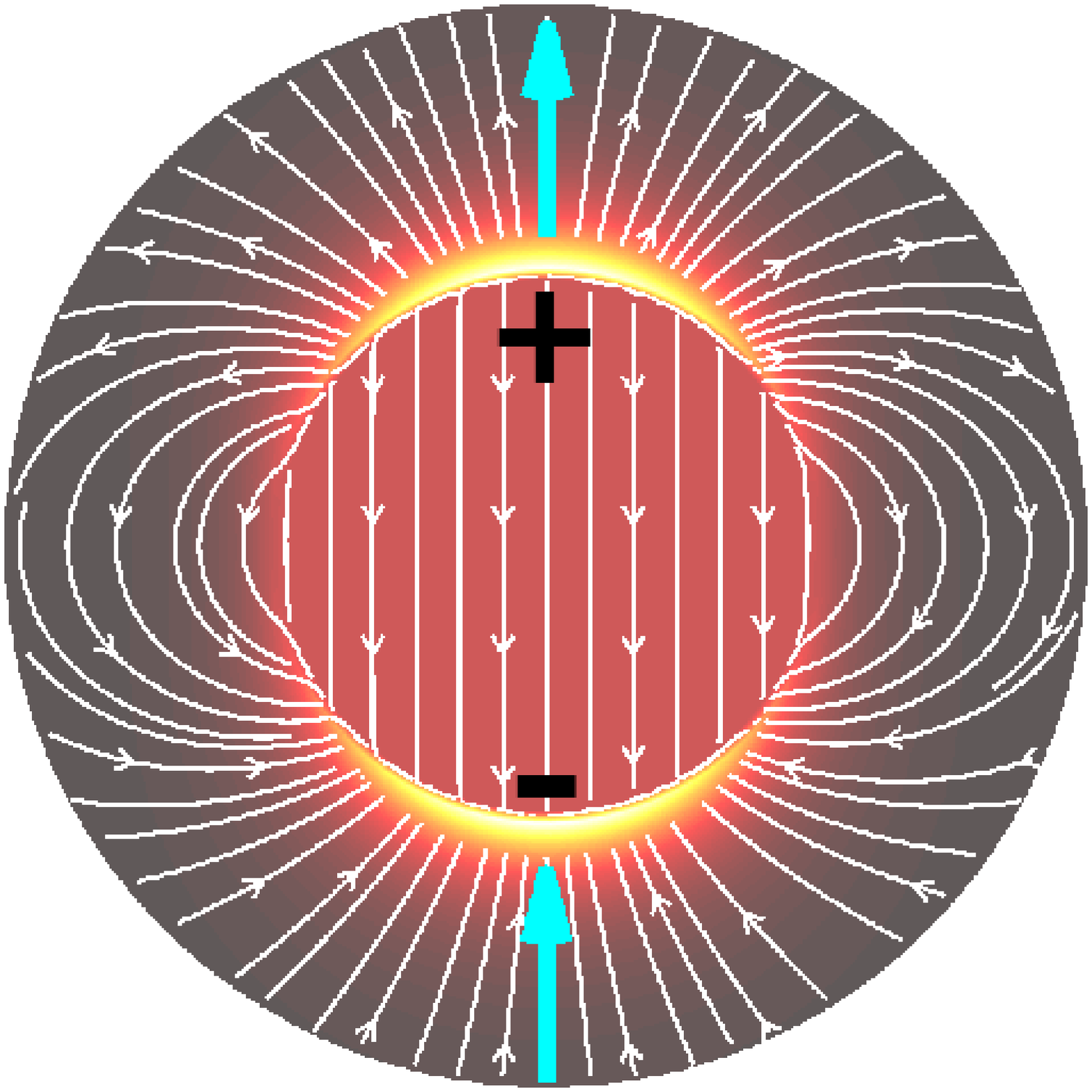} &
 \includegraphics[width=2.25cm]{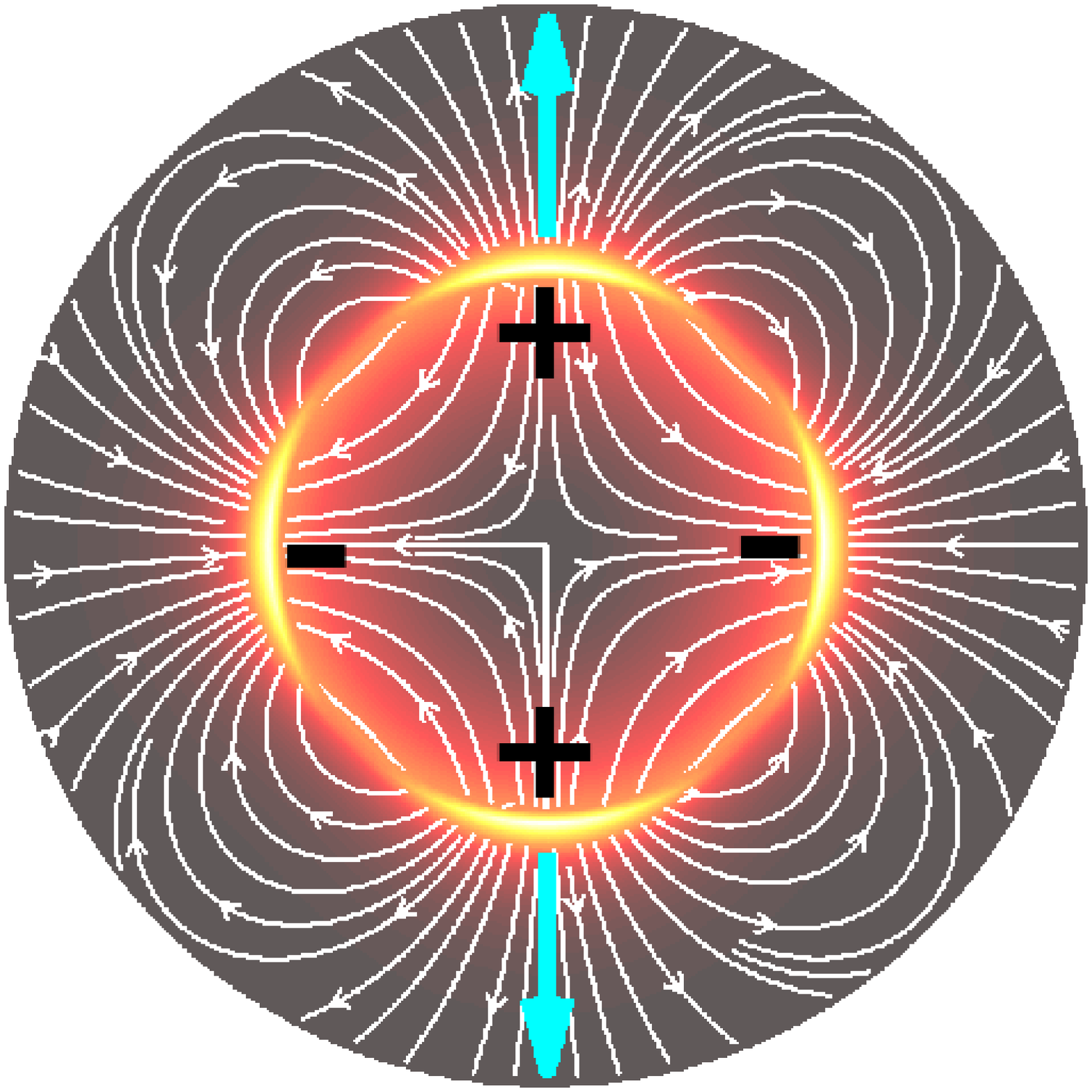} &
\includegraphics[width=2.25cm]{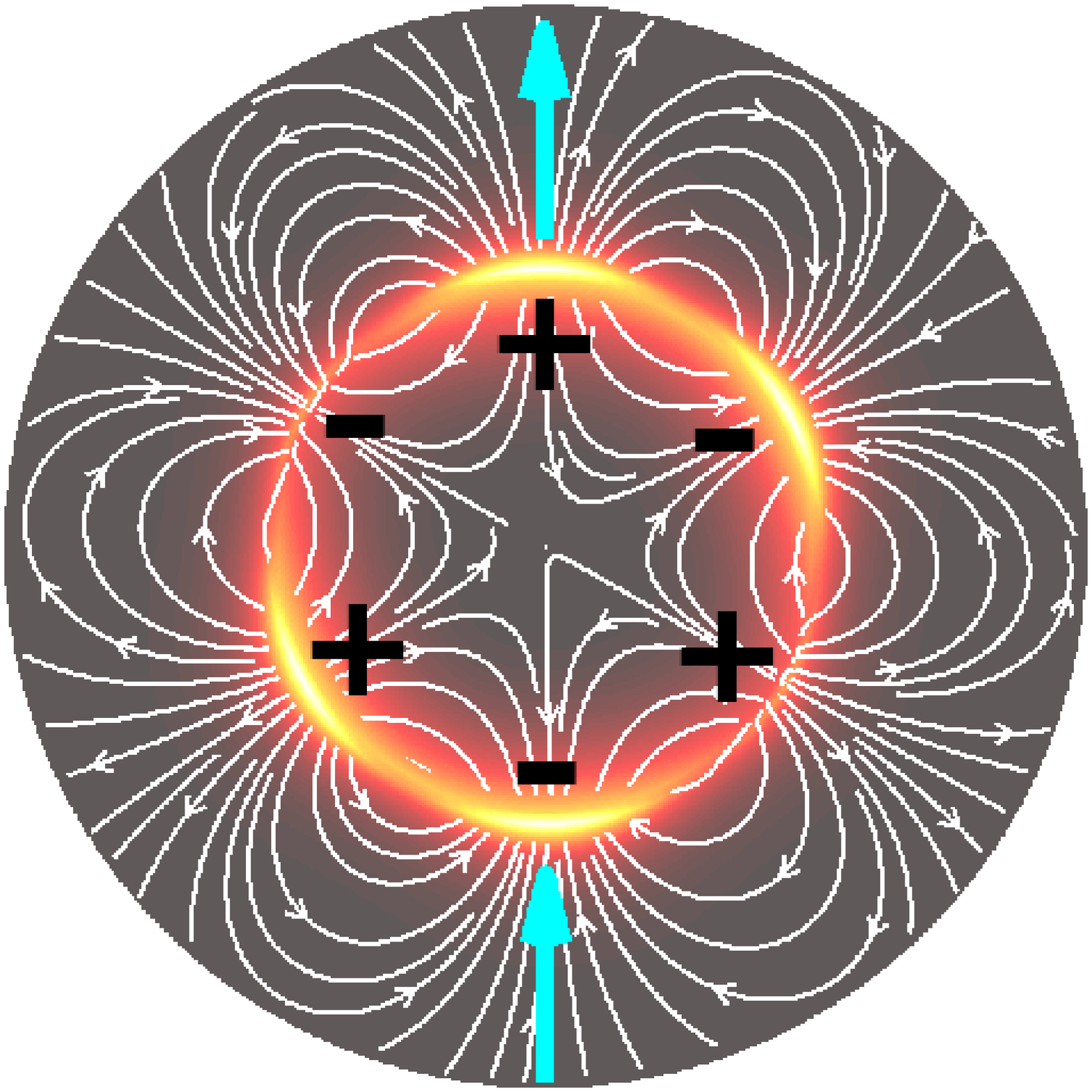}  \\
\hline 
LSP$_n$ & n=1 & n=2 & n=3
\end{tabular}
\caption{Plasmonic bright states considering single mode MNP response (LSP$_n$) and 2 radial emitters ($h=2$ nm, $\omega_0=2.964$ eV). The mode field lines and charge density are indicated.}  
\label{table:DickeLSPn}
\end{table}
The superradiance blockade can be also inferred from the eigenvalue model (Eq. \ref{eq:ClassicSuper}). The eigenvectors are represented in table \ref{table:Dicke2}. We check that $\Gamma_S+\Gamma_A=2\Gamma_1$ for all symmetric/antisymmetric pairs in agreement with the Fig. \ref{scheme12}. However, for small separation distance $h=2$ nm and $\omega_0=2.964$ eV, corresponding to high order modes resonance, the cooperative rate vanishes for both radial or azimuthal emitters and the dynamics is almost incoherent. In order to understand this behaviour more deeply, we consider each mode separately on table \ref{table:DickeLSPn}. The alternating charge distribution at the position of the dipole clearly leads to cancelation of the cooperative behaviour by destructive interferences. This corresponds to a coupling strength between the two emitters $\mu_{odd}^{12}\approx 1$, $\mu_{even}^{12}\approx -1$. Therefore, the quantum collective decay rate $\Gamma_{12}$ vanishes at short distances when numerous LSPs modes are involved (Eq. \ref{pertes}). It is worth noticing that the collective rate plays a role in the whole quantum superradiance cascade dynamics and not only for the single excitation state, which is the only one considered in the eigenvalue model. This explains the discrepancy between the partial superradiance observed for radial emitters in the quantum approach (with 6 excitations as initial state, Fig. \ref{It}b) and the almost zero decay rate of the single excitation state (see table \ref{table:Dicke6All}), that is not populated during the superradiance process.   

\paragraph*{Conclusion and outlook.} We have derived a quantum approach for plasmonic superradiance and discussed the dynamics of cooperative emission. The system follows the plasmonic Dicke ladder and strong  LSP mediated correlations build up between the emitters so that superradiance can be Purcell enhanced. Since the build-up of the cooperative emission ($\sim 1/N_e \Gamma_1$) has to be smaller than the dephasing rate $\gamma^\star$ \cite{Gross-Haroche:1982}, this would facilitate the conditions to achieve superradiance.  However, superradiance blockade occurs at small distances where cooperativity is jeopardized because of modal destructive interferences. Finally, our work brings deeper understanding of light-matter interaction at the nanoscale which can be helpful in designing surface plasmon laser (SPASER) since it relies on a very similar mechanism \cite{Bergman2003}. We considered identical emitters throughout this work and we expect that the superradiance should be degraded for non identical emitters. However, some recent works indicate that cooperative dissipation in a nanocrystal could help to recover the superradiance effect \cite{JuanPRL:18} and we expect that plasmonic superradiance could be experimentally investigated on metallo-dielectric nanohybrids \cite{Noginov2009,Zhou-GCF-Bachelot:2015,Greffet-Dubertret:2015,FauchePhD:17}. More generally, our approach opens up the possibility to systematically optimize the superradiance effect for different geometries of the nanoparticles \cite{Franke-Hughes:18}.

\begin{acknowledgments}
This work is supported by the French "Investissements d'Avenir" program: ISITE-BFC IQUINS (ANR-15-IDEX-03 ) and EUR-EIPHI (17-EURE-0002). This work is part of the  european COST action MP1403 Nanoscale Quantum Optics. Additional support from the European Union's Horizon 2020 research and innovation program under the Marie Sklodowska-Curie grant agreement No. 765075 (LIMQUET) is acknowledged.
\end{acknowledgments}


\newpage
~   
\newpage

\onecolumngrid
\renewcommand*{\citenumfont}[1]{S#1}
\renewcommand*{\bibnumfmt}[1]{[S#1]}

\center
{\bf Cooperative emission by quantum plasmonic superradiance: \\Supplemental material}

\section{Effective Hamiltonian}
\subsection{Multi-emitters Hamiltonian}
We consider $N_e$ identical emitters located at a distance $h$ from a metal nanoparticle (MNP) of radius $R$. The emitters consists of two-level systems (TLS). The optical transition between the excited state $\ket{e}$ and ground state $\ket{g}$ is characterized by the angular frequency $\omega_0$ and the transition dipole moment $\mathbf{d}^{(i)}=d_0\mathbf{e}_i$ where $\mathbf{e}_i$ is a unit vector corresponding to the orientation of the  dipole $i$. The effective Hamiltonian that governs the dynamics of the system is \cite{Dzsotjan-GCF:2016}
\begin{eqnarray}
\hat{H}=&\sum_{n=1}^{N}\hbar(\delta_n-i\gamma_n)\sum_{i=1}^{N_e}\hat{a}_n^{(i)\dagger}\hat{a}^{(i)}_n \nonumber\\
&-i\hbar\sum_{n=1}^{N}\sum_{i=1}^{N_e} \left(g^{(i)}_n\hat{\sigma}^{(i)}_+\hat{a}^{(i)}_n-(g_n^{(i)})^*\hat{a}_n^{(i)\dagger}\hat{\sigma}^{(i)}_-\right)
,\label{H_Ne}
\end{eqnarray}
where  we omit the free-space losses of the emitters since they are negligibly small compared to the LSP losses. We introduced the detuning factor $\delta_n=\omega_n-\omega_0$ and the operators $\hat{\sigma}_+^{(i)}$ and $\hat{\sigma}_-^{(i)}$ that are the transition operators associated to emitter $i$, respectively. The bosonic operators $\hat{a}_n^{(i)\dagger}$ and $\hat{a}_n^{(i)}$ describe the creation or annihilation of a plasmon excitation of order $n$ associated to emitter $i$. The model is derived from an quantization of the LSPs. It leads to a coupling constant between emitter ($i$) and LSP$_n$ that displays a Lorentzian profile, 
\begin{eqnarray}
\kappa_n^{(i)}(\omega)=\sqrt{\frac{\gamma_n}{2\pi}}\frac{g_n^{(i)}}{\omega-\omega_n+i\frac{\gamma_n}{2}} \;, \label{kappa_complexe}
\end{eqnarray}
with the coupling strength $g_n^{(i)}$ to LSP$_n$ and the LSP resonance angular frequency $\omega_n$ and width $\gamma_n$.
It is worthwhile to note that the $\hat{a}^{(i)}_n$ operators associated to different emitters do not obey the standard bosonic commutation relations: 
\begin{eqnarray}
&&\left[\hat{a}_n^{(j)},\hat{a}_{n }^{(k)\dagger}\right]=\delta_{nn }\mu_n^{(jk)},\label{com} \\
&&\mu_n^{(jk)}=\frac{1}{\pi \hbar\varepsilon_0}\frac{\omega^2}{c^2}\frac{ \mathfrak{Im}\left[\mathbf{d}^{(j)}\cdot\mathbf{G_n}(\mathbf{r}_j,\mathbf{r}_k,\omega)\cdot \mathbf{d}^{(k)}\right]}{\kappa_n^{(j)}(\omega)\kappa_n^{(k)^*}(\omega)} \;,
\label{eq:muij}
\end{eqnarray}
where $\mu_n^{(jk)}$ is the modal overlap function. 
One can define operators satisfying the standard commutation rules by a L\"owdin orthormalisation \cite{Annavaparu:13,Castellini:18}. We first define the overlap matrix $\mathbf{M}_n$ with the overlap functions for a given order $n$
\begin{eqnarray}
\mathbf{M}_n=\begin{pmatrix}
1 & \mu_n^{(21)} & \cdots & \mu_n^{(N_e1)}\\
\mu_n^{(12)} & 1 & \cdots & \mu_n^{(N_e2)}\\
\vdots & \vdots & \ddots & \vdots\\
\mu_n^{(1N_e)} & \mu_n^{(2N_e)} & \cdots & 1\end{pmatrix}.
\end{eqnarray}
The new operators $\hat{l}_n^{(j)}$ are
\begin{eqnarray}
\hat{l}_n^{(j)} & =\lambda_{j,n}^{-1/2}\sum_{i=1}^{N_e}T^{ij}_n\hat{a}_n^{(i)},\label{low}\\
\hat{a}_n^{(i)} & =\sum_{j=1}^{N_e}\lambda_{j,n}^{1/2}\left(T_n^{ij}\right)^*\hat{l}_n^{(j)}\label{low2},
\end{eqnarray}
where $\lambda_{j,n}$ are the eigenvalues of $\mathbf{M}_n$ and $\mathbf{T}_n$ the matrix diagonalising $\mathbf{M}_n$. The creation and annihilation operators $\hat{l}_n^{(i)\dagger}$ and $\hat{l}^{(i)}_n$ constitute an orthormal basis of $N_{ind}$ operators associated to the LSP$_n$ modes and satisfy the commutation relations
\begin{eqnarray}
\left[\hat{l}_{n}^{(i)},\hat{l}_{n'}^{(j)\dagger}\right]&=&\delta_{nn'}\delta_{ij},\\
\left[\hat{l}_{n}^{(i)},\hat{l}_{n'}^{(j)}\right]&=&0.
\end{eqnarray}

By inserting Eq. \eqref{low2} into the Hamiltonian \eqref{H_Ne} we derive \cite{Castellini:18}
\begin{eqnarray}
\hat{H}_{eff}=\sum_{n=1}^{N}\hbar\left(\delta_n-i\frac{\gamma_n}{2}\right)\sum_{i=1}^{N_{ind}}\hat{l}_n^{(i)\dagger}\hat{l}^{(i)}_n
& - i\hbar\sum_{n=1}^{N}\sum_{j=1}^{N_e}\sum_{i=1}^{N_{ind}} \left(g^{(ji)}_n\hat{\sigma}^{(j)}_+\hat{l}^{(i)}_n-\left(g_n^{(ji)}\right)^*\hat{l}_n^{(i)\dagger}\hat{\sigma}^{(j)}_-\right) 
\label{Heff}
\end{eqnarray}
with $g_n^{(ij)}=g_n^{(i)}\lambda_{j,n}^{1/2}\left(T^{ij}_n\right)^*$ the cross-coupling between the emitter $i$ and the L\"owdin mode $j$.
\subsection{Reformulation of the dissipation}
The expression (\ref{Heff}) is fully analogous to cQED description for a lossy multimodal cavity. By analogy to the cQED treatment, we can describe LSPs dissipation by the coupling to a continuum bath. We separate the system $\mathcal{S}$ made of emitters and LSPs from the environment $\mathcal{E}$ associated to the bath. Hence, we introduce a new Hamiltonian leading to the same dynamics as the original one

\begin{eqnarray}
\hat{H}_{SE}&=&H_S+H_E+H_I,\\
H_S&=&\sum_{n=1}^{N}\hbar\delta_n\sum_{i=1}^{N_{ind}}\hat{l}_n^{(i)\dagger}\hat{l}^{(i)}_n- i\hbar\sum_{n=1}^{N}\sum_{j=1}^{N_e}\sum_{i=1}^{N_{ind}} \left(g^{(ji)}_n\hat{\sigma}^{(j)}_+\hat{l}^{(i)}_n-\left(g_n^{(ji)}\right)^*\hat{l}_n^{(i)\dagger}\hat{\sigma}^{(j)}_-\right),\label{H_S}\\
H_E&=&\int d\omega\,\hbar\omega\sum_{n=1}^{N}\sum_{i=1}^{N_{ind}}\hat{b}_{\omega,n}^{(i)\dagger}\hat{b}^{(i)}_{\omega,n},\\
H_{I}&=&i\hbar\int d\omega\,\sum_{n=1}^{N}\sum_{i=1}^{N_{ind}}\beta_n^{(i)}(\omega)\left(\hat{b}_{\omega,n}^{(i)\dagger}\hat{l}_n^{(i)}-\hat{l}_{n}^{(i)\dagger}\hat{b}^{(i)}_{\omega,n}\right).\label{H_I}
\end{eqnarray}
The system Hamiltonian $H_S$ describes the interaction between the emitters and the LSPs and is hermitian. The environment Hamiltonian $H_E$ involves all the bath oscillators of energy $\hbar\omega$. For each cavity pseudo-mode of order $n$ and index $i$ associated to the operators $\hat{l}_n^{(i)\dagger}$ and $\hat{l}^{(i)}_n$, we define a reservoir. The creation  and annihilation operators $\hat{b}^{(i)\dagger}_{\omega,n}$ and $\hat{b}^{(i)}_{\omega,n}$, associated to these reservoirs, satisy the commutations relations 
\begin{eqnarray}
\left[\hat{b}^{(i)}_{\omega,n},\hat{b}_{\omega',n'}^{(j)\dagger}\right]&=&\delta_{nn'}\delta(\omega-\omega')\delta_{ij},\\
\left[\hat{b}^{(i)}_{\omega,n},\hat{b}^{(j)}_{\omega',n'}\right]&=&0.
\end{eqnarray}
The interaction between the system and the environment is described by the Hamiltonian $H_I$ where $\beta_n^{(i)}(\omega)$ characterizes the coupling between the pseudo-modes and their associated reservoirs.

\section{Heisenberg representation and Langevin equation}
The dynamics of the system is investigated using the Heisenberg-Langevin equation. For any operator $\hat{O}$ in the Schr{\"o}dinger representation, its Heisenberg representation is 
\begin{eqnarray}
\hat{O}(t)=\hat{U}^{\dagger}(t,t_0)\hat{O}\hat{U}(t,t_0) \;,
\label{heis}
\end{eqnarray} 
where $\hat{U}(t,t_0)$ is the time propagator from time $t_0$ to $t$. In the following, the explicit time dependence refers to the Heisenberg representation. The dynamics follows 
\begin{eqnarray}
\frac{\partial}{\partial t}\hat{O}(t)=-\frac{i}{\hbar}\left[\hat{O}(t),\hat{H}_{SE}(t)\right]. 
\end{eqnarray}
We first consider the Langevin equation for the bath operators 
\begin{eqnarray}
\frac{\partial}{\partial t}\left(\hat{b}^{(i)}_{\omega,n}(t)\right)=-i\omega\hat{b}^{(i)}_{\omega,n}(t)+\beta_n^{(i)}(\omega)\hat{l}_n^{(i)}(t).\label{diff_b}
\end{eqnarray}
Following standard procedures (see {\it e.g} \S 5.3 in ref. \cite{Gardiner-Zoller:04}), and notably within the Markov approximation by assuming a flat coupling between the LSPs and the free-space continuum [$\beta_n^{(i)}(\omega)\approx\beta_n^{(i)}(\omega_n)=\left(\gamma_n/2\pi\right)^{1/2}$], we can derive 
\begin{eqnarray}
\hat{b}_n^{(i)}(t)=\hat{b}^{(i)}_{\varnothing,n}(t)+\frac{\sqrt{\gamma_n}}{2}\hat{l}_n^{(i)}(t) \;,\label{bint_final}
\end{eqnarray}
where we have defined the effective operators 
\begin{eqnarray}
\hat{b}^{(i)}_n(t)&=&\frac{1}{\sqrt{2\pi}}\int d\omega\, \hat{b}^{(i)}_{\omega,n}(t) \label{def_bint} \;,\\
\hat{b}^{(i)}_{\varnothing,n}(t)&=&\frac{1}{\sqrt{2\pi}}\int d\omega\,\hat{b}^{(i)}_{\omega,n}\e^{-i\omega(t-t_0)} \;.\label{def_b0int}
\end{eqnarray}
$\hat{b}^{(i)}_{\varnothing,n}(t)$ is the isolated effective bath operator and does not interact with the system. 
We can now derive the dynamics of any system operator $\hat{X}_S(t)$ 
\begin{eqnarray}
\frac{\partial}{\partial t}\left(\hat{X}_S(t)\right)&=&-\frac{i}{\hbar}\left[\hat{X}_S(t),\hat{H}_{S}(t)\right]
+\sum_{n=1}^{N}\sum_{i=1}^{N_{ind}}\gamma_n\left(\hat{l}_n^{(i)\dagger}(t)\hat{X}_S(t)\hat{l}_n^{(i)}(t)-\frac{1}{2}\left\{\hat{l}_n^{(i)\dagger}(t)\hat{l}_n^{(i)}(t),\hat{X}_S(t)\right\}\right)\nonumber\\
&+&\sum_{n=1}^{N}\sum_{i=1}^{N_{ind}}\sqrt{\gamma_n}\left(\hat{b}_{\varnothing,n}^{(i)\dagger}(t)\left[\hat{X}_S(t),\hat{l}_n^{(i)}(t)\right]-\left[\hat{X}_S(t),\hat{l}_n^{(i)\dagger}(t)\right]\hat{b}_{\varnothing,n}^{(i)}(t)\right)
,\label{deriv_Xs}
\end{eqnarray}
where $\left\{ ~,~\right\}$ denotes the anti-commutator.

\section{Dissipative regime and adiabatic elimination}
The numerical simulation of the dynamics of the system can become heavy for a large number of emitters and LSPs. Since superradiance is a strongly dissipative phenomenum, we restrict our study to the weak coupling regime
\begin{eqnarray}
\sqrt{\sum_{n=1}^N \frac{\sum_{i=1}^{N_e}\left(g^{(i)}_n\right)^2}{\delta_n^2+\left(\frac{\gamma_n}{2}\right)^2}} \ll 1 \;.\label{ratio}
\end{eqnarray}
This condition generalizes the weak coupling condition to several emitters and several modes. It is then possible to adiabatically eliminate the plasmon modes and to work on the emitter states basis only. 
The LSPs are practically not populated due to their strong dissipation. The adiabatic elimination therefore permits to report LSPs losses to the excited emitter. Specifically, we assume that the wavefunction evolves in the Hilbert subpsace associated to the emitter and environment only (that is without the LSPs) so that

\begin{eqnarray}
\frac{\partial\,\hat{l}_n^{(i)}(t)}{\partial t}&\approx 0 \;, 
\end{eqnarray}
which is a concise way to derive the effective Hamiltonian after adiabatic elimination \cite{Vitanov-Stenholm:97}. Using the Heisenberg-Langevin equation \eqref{deriv_Xs} for the operators $\hat{l}_n^{(i)}(t)$, we obtain
\begin{eqnarray}
\hat{l}_n^{(i)}(t)&\approx&i\frac{\sqrt{\gamma_n}\hat{b}_{\varnothing,n}^{(i)}(t)-\sum_{j=1}^{N_e}\left(g_n^{(ji)}\right)^*\sigma_-^{(j)}(t)}{\delta_n-i\frac{\gamma_n}{2}} \;.\label{l_eli}
\end{eqnarray}

Finally, the dynamics of any system operator is derived from the Heisenberg-Langevin equation (\ref{deriv_Xs}), using the adiabatic approximation expression for the operators $\hat{l}_n^{(i)}(t)$ and $\hat{l}_n^{(i)\dagger}(t)$. This leads to
\begin{eqnarray}
\frac{\partial}{\partial t}\left(\hat{X}_S(t)\right)&=i\sum_{j=1}^{N_e}\sum_{k=1}^{N_e}\Delta_{jk}\left[\hat{X}_S(t),\hat{\sigma}_+^{(k)}(t)\hat{\sigma}_-^{(j)}(t)\right]-\frac{i}{\hbar}\sum_{n=1}^N\sum_{i=1}^{N_{ind}}\left[\hat{X}_S(t),\hat{H}_{\varnothing,n}^{(i)}(t)\right]\nonumber\\
+&\sum_{n=1}^{N}\sum_{i=1}^{N_{ind}}\hat{\mathcal{D}}_{\varnothing,n}^{(i)}\left(\hat{X}_S(t)\right)+\sum_{j=1}^{N_e}\sum_{k=1}^{N_e}\Gamma_{jk}\left[\hat{\sigma}_+^{(k)}(t)\hat{X}_S(t)\hat{\sigma}_-^{(j)}(t)-\frac{1}{2}\left\{\hat{\sigma}_+^{(k)}(t)\hat{\sigma}_-^{(j)}(t),\hat{X}_S(t)\right\}\right]\label{kjl},
\end{eqnarray}
where we have introduced the collective decay rates $\Gamma_{jk}$ and Lamb shift $\Delta_{jk}$
\begin{eqnarray}
\Gamma_{jk}&=&\sum_{n=1}^{N}\frac{\gamma_n}{\delta_n^2+\left(\frac{\gamma_n}{2}\right)^2}\sum_{i=1}^{N_{ind}} \left(g_n^{(ji)}\right)^*g_n^{(ki)} \\
\Delta_{jk}&=&\sum_{n=1}^{N}\frac{\delta_n}{\delta_n^2+\left(\frac{\gamma_n}{2}\right)^2}\sum_{i=1}^{N_{ind}} \left(g_n^{(ji)}\right)^*g_n^{(ki)} \;,
\end{eqnarray}
and $\mathcal{D}_{\varnothing,n}^{(i)}\left(\hat{X}_S(t)\right)$  and $\hat{H}_{\varnothing,n}^{(i)}(t)$ involve the isolated effective bath operator $\hat{b}_{\varnothing,n}^{(i)}(t)$.  

Note that using the relation $g_n^{(ij)}=g_n^{(i)}\lambda_{j,n}^{1/2}\left(T^{ij}_n\right)^*$ and Eq. (\ref{eq:muij}), we can write equivalently 
\begin{eqnarray}
\Gamma_{jk}&=& \sum_{n=1}^{N}\frac{\gamma_n}{\delta_n^2+\left(\frac{\gamma_n}{2}\right)^2} g_n^{(j)}g_n^{(k)}\mu_n^{(jk)} \\
&=&\frac{2\omega_0^2}{\hbar\varepsilon_0 c^2}\mathfrak{Im}\left[\mathbf{d}^{(j)}\cdot \mathbf{G}(\mathbf{r}_j,\mathbf{r}_k,\omega_0)\cdot \mathbf{d}^{(k)}\right]\;.
\end{eqnarray}

Equation (\ref{kjl}) applies to any system operator that does not involve the cavity like operators $\hat{l}_n^{(i)}(t)$, since they have been adiabatically eliminated. In the following, we assume that the system involves only the emitters; $\mathcal{S} \equiv \mathcal{A}t$ where $\mathcal{A}t$ refers to the  atomic (emitters)  subspace (no LSP).

\section{Master equation}
Let us introduce the density operator $\hat{\rho}(t)$ defined on $\mathcal{A}t\oplus\mathcal{E}$. The operator average is 
\begin{eqnarray}
\left\langle \hat{X}_{At} \right\rangle(t)=\text{Tr}\left\{\hat{X}_{At}\hat{\rho}(t)\right\}=\text{Tr}\left\{\hat{X}_{At}(t)\hat{\rho}(t_0)\right\},\label{moy_Xs}
\end{eqnarray}
We assume the initial decomposition $\hat{\rho}(t_0)=\hat{\rho}_{At}(t_0)\otimes\hat{\rho}_E(t_0)$ 
The derivation of eq. (\ref{moy_Xs}) leads to 
\begin{eqnarray}
\frac{\partial}{\partial t}\left\langle \hat{X}_{At} \right\rangle(t)&=&\text{Tr}\left\{\frac{\partial\hat{X}_{At}(t)}{\partial t} \hat{\rho}(t_0)\right\} \\
 \nonumber
&=&i\sum_{j=1}^{N_e}\sum_{k=1}^{N_e}\Delta_{jk}\text{Tr}\left\{\left[\hat{X}_{At}(t),\hat{\sigma}_+^{(k)}(t)\hat{\sigma}_-^{(j)}(t)\right]\hat{\rho}(t_0)\right\} \nonumber\\
&-&\frac{i}{\hbar}\sum_{n=1}^N\sum_{i=1}^{N_{ind}}\text{Tr}\left\{\left[\hat{X}_{At}(t),\hat{H}_{\varnothing,n}^{(i)}(t)\right]\hat{\rho}(t_0)\right\}+\sum_{n=1}^{N}\sum_{i=1}^{N_{ind}}\text{Tr}\left\{\hat{\mathcal{D}}_{\varnothing,n}^{(i)}\left(\hat{X}_{At}(t)\right)\hat{\rho}(t_0)\right\}\nonumber\\
&+&\sum_{j=1}^{N_e}\sum_{k=1}^{N_e}\Gamma_{jk}\left(\text{Tr}\left\{\hat{\sigma}_+^{(k)}(t)\hat{X}_{At}(t)\hat{\sigma}_-^{(j)}(t)\hat{\rho}(t_0)\right\}\right)\nonumber\\
&-&\sum_{j=1}^{N_e}\sum_{k=1}^{N_e}\frac{\Gamma_{jk}}{2}\left(\text{Tr}\left\{\hat{\sigma}_+^{(k)}(t)\hat{\sigma}_-^{(j)}(t)\hat{X}_{At}(t)\hat{\rho}(t_0)\right\}+\text{Tr}\left\{\hat{X}_{At}(t)\hat{\sigma}_+^{(k)}(t)\hat{\sigma}_-^{(j)}(t)\hat{\rho}(t_0)\right\}\right).\label{deriv_Xs_2}
\end{eqnarray}
The terms $\left[\hat{X}_{At}(t),\hat{H}_{\varnothing,n}^{(i)}(t)\right]$ and $\hat{\mathcal{D}}_{\varnothing,n}^{(i)}\left(\hat{X}_S(t)\right)$ simplify taking into account the initial condition $\hat{\rho}_E(t_0)=\vert\varnothing_E\rangle\langle\varnothing_E\vert$. Indeed, from the definition (\ref{def_b0int}), the effective bath operators $\hat{b}_{\varnothing,n}^{(i)}(t)$ acts on the environment subspace $\mathcal{E}$ only so that they commute with $\hat{\rho}_{At}(t)$ . Therefore $\hat{b}_{\varnothing,n}^{(i)}(t)\hat{\rho}(t_0)=0$. 
We finally obtain 
\begin{eqnarray}
\text{Tr}&\left\{\frac{\partial\hat{X}_{At}(t)}{\partial t} \hat{\rho}(t_0)\right\}=i\sum_{j=1}^{N_e}\sum_{k=1}^{N_e}\Delta_{jk}\left(\text{Tr}\left\{\hat{X}_{At}(t)\hat{\sigma}_+^{(k)}(t)\hat{\sigma}_-^{(j)}(t)\hat{\rho}(t_0)\right\}-\text{Tr}\left\{\hat{\sigma}_+^{(k)}(t)\hat{\sigma}_-^{(j)}(t)\hat{X}_{At}(t)\hat{\rho}(t_0)\right\}\right)\nonumber\\
&+\sum_{j=1}^{N_e}\sum_{k=1}^{N_e}\Gamma_{jk}\left(\text{Tr}\left\{\hat{\sigma}_+^{(k)}(t)\hat{X}_{At}(t)\hat{\sigma}_-^{(j)}(t)\hat{\rho}(t_0)\right\}\right)\nonumber\\
&-\sum_{j=1}^{N_e}\sum_{k=1}^{N_e}\frac{\Gamma_{jk}}{2}\left(\text{Tr}\left\{\hat{\sigma}_+^{(k)}(t)\hat{\sigma}_-^{(j)}(t)\hat{X}_{At}(t)\hat{\rho}(t_0)\right\}+\text{Tr}\left\{\hat{X}_{At}(t)\hat{\sigma}_+^{(k)}(t)\hat{\sigma}_-^{(j)}(t)\hat{\rho}(t_0)\right\}\right) \;.
\end{eqnarray}
Then, we work in the Schr{\"o}dinger representation using relation (\ref{heis}) and taking benefit of trace conservation by circular permutation ($\text{Tr}\left\{ABC\right\}=\text{Tr}\left\{BCA\right\}=\text{Tr}\left\{CAB\right\}$). We obtain 
\begin{eqnarray} 
\text{Tr}\left\{\frac{\partial\hat{X}_{At}(t)}{\partial t} \hat{\rho}(t_0)\right\}&=&\text{Tr}_{At}\left\{\hat{X}_{At}\left[\sum_{j=1}^{N_e}\sum_{k=1}^{N_e}-\frac{i}{\hbar}\left[\hat{H}_{jk},\hat{\rho}_{At}(t)\right]+\Gamma_{jk}\left(\hat{\sigma}_-^{(j)}\hat{\rho}_{At}(t)\hat{\sigma}_+^{(k)}-\frac{1}{2}\left\{\hat{\sigma}_+^{(k)}\hat{\sigma}_-^{(j)},\hat{\rho}_{At}(t)\right\}\right)\right]\right\} \;,
\end{eqnarray}
with $\hat{H}_{jk}=-\hbar\Delta_{jk}\hat{\sigma}_+^{(k)}\hat{\sigma}_-^{(j)}$. Finally, using the equality (\ref{moy_Xs}), it comes 
\begin{eqnarray}
&\text{Tr}\left\{\hat{X}_S\frac{\partial\hat{\rho}(t)}{\partial t}\right\}=\text{Tr}_S\left\{\hat{X}_S\frac{\partial\hat{\rho}_S(t)}{\partial t}\right\}\nonumber\\
&=\text{Tr}_S\left\{\hat{X}_S\left[\sum_{j=1}^{N_e}\sum_{k=1}^{N_e}-\frac{i}{\hbar}\left[\hat{H}_{jk},\hat{\rho}_S(t)\right]+\Gamma_{jk}\left(\hat{\sigma}_-^{(j)}\hat{\rho}_S(t)\hat{\sigma}_+^{(k)}-\frac{1}{2}\left\{\hat{\sigma}_+^{(k)}\hat{\sigma}_-^{(j)},\hat{\rho}_S(t)\right\}\right)\right]\right\} \;.\label{presque}
\end{eqnarray}
Last, the master equation is obtained from relation (\ref{presque})
\begin{eqnarray}
\frac{\partial\hat{\rho}_S(t)}{\partial t}=\sum_{j=1}^{N_e}\sum_{k=1}^{N_e}-\frac{i}{\hbar}\left[\hat{H}_{jk},\hat{\rho}_S(t)\right]+\Gamma_{jk}\left(\hat{\sigma}_-^{(j)}\hat{\rho}_S(t)\hat{\sigma}_+^{(k)}-\frac{1}{2}\left\{\hat{\sigma}_+^{(k)}\hat{\sigma}_-^{(j)},\hat{\rho}_S(t)\right\}\right). \label{eq_maitresse}
\end{eqnarray}
since $\text{Tr}\left\{AB\right\}=\text{Tr}\left\{AC\right\}$ for all $A$ implies $B=C$. This master equation is in Lindblad form (see ref. \cite{Breuer-Petruccione:02} p. 121).

We note that if $\Gamma_{jk}=\pm \sqrt{\Gamma_{j}\Gamma_{k}}$, the dissipator is in the diagonal Lindblad form 
\begin{eqnarray}
\mathcal{D}_{jk}\left[\hat{\rho}(t)\right]&=&\left[\hat{c}_-^{(jk)}\hat{\rho}(t)\hat{c}_+^{(jk)} \right. 
\label{Djk0} \\
&& \left. - \frac{1}{2}\left(\hat{c}_+^{(jk)}\hat{c}_-^{(jk)}\hat{\rho}(t)+\hat{\rho}(t)\hat{c}_+^{(jk)}\hat{c}_-^{(jk)}\right) \right] \,.
\nonumber
\end{eqnarray}
where we define the collapse operator $\hat{c}_+^{(jk)}=\sqrt{\Gamma_j}\hat{\sigma}_{-}^{(j)}\pm\sqrt{\Gamma_k}\hat{\sigma}_{-}^{(k)}$. Then, the single excitation bright state is 
$\vert B\rangle=\sqrt{\Gamma_1}\vert eg\dots g\rangle+s_2\sqrt{\Gamma_2}\vert ge \dots g\rangle 
+s_{N_e}\sqrt{\Gamma_{N_e}}\vert g\dots ge\rangle$ ($s_j=\Gamma_{1j}/\vert\Gamma_{1j}\vert$) with the maximized decay rate $\Gamma_{tot}=\sum_{i=1}^{N_e}\Gamma_i$ corresponding to highly correlated emitters close to the ideal superradiance situation.

\section{Vector representation of the density operator}
The vector representation of the density operator is used in the main text to compute the decay rates along the extended Dicke cascade. We reproduce the expression, taken from Ref. \cite{NavarreteBenlloch:15} for the sake of completeness.  
If the density operator is
\begin{eqnarray}
\hat \rho= \sum_{n,m}\rho_{nm}\vert n\rangle \langle m\vert
\end{eqnarray}
then, it can be numerically manipulated considering the vector representation 
\begin{eqnarray}
\vert \widetilde \rho ~ \rangle= \sum_{n,m}\rho_{nm}\vert n\rangle \otimes  \vert m \rangle,.
\end{eqnarray}
Similarly, the vector representation of the projector $\vert  J,M\rangle \langle   J,M \vert$ on the Dicke state $\vert  J,M\rangle$ is 
$\vert \widetilde{M}\rangle= \vert  J,M\rangle \otimes \vert  J,M\rangle $.

The dissipator of the system reads 
\begin{eqnarray}
\mathcal{D}\left[\hat{\rho}(t)\right]=\sum_{j,k=1}^{N_e}\mathcal{D}_{jk}\left[\hat{\rho}(t)\right]=\sum_{j=1,k}^{N_e}\Gamma_{jk}\left[\hat{\sigma}_-^{(j)}\hat{\rho}(t)\hat{\sigma}_+^{(k)}  - \frac{1}{2}\left(\hat{\sigma}_+^{(k)}\hat{\sigma}_-^{(j)}\hat{\rho}(t)+\hat{\rho}(t)\hat{\sigma}_+^{(k)}\hat{\sigma}_-^{(j)}\right) \right] 
\end{eqnarray} 
and the form associated to the vector representation of the density operator is
\begin{eqnarray}
\mathcal{D}_{\vert \rho\rangle}=\sum_{j=1,k}^{N_e}\Gamma_{jk}\left[\hat{\sigma}_-^{(j)}\otimes (\hat{\sigma}_-^{(k)})^\star  - \frac{1}{2}\hat{\sigma}_+^{(k)}\hat{\sigma}_-^{(j)}\otimes   \mathbb{1}  - \frac{1}{2}  \mathbb{1}\otimes  (\hat{\sigma}_-^{(j)})^T(\hat{\sigma}_-^{(k)})^\star \right] \,.
\end{eqnarray}


\end{document}